\documentclass[aps,prb,showpacs,twocolumn,superscriptaddress]{revtex4} 

\usepackage[]{graphicx,color}% Include figure files
\usepackage[version=3]{mhchem}

\usepackage{ulem}

\begin{document}

\title{Evolutionary search for cobalt-rich compounds in the yttrium-cobalt-boron system
%Prediction of novel \ce{YFe_{12}} phases from first principles
%Novel monoclinic phases in \ce{YFe_{12}} predicted from first principles
%Monoclinic \ce{YFe_{12}} phases predicted from first principles
}



\author{Takahiro Ishikawa}%
 \email{ISHIKAWA.Takahiro@nims.go.jp}
 \affiliation{%
 ESICMM, National Institute for Materials Science, 1-2-1 Sengen, Tsukuba, Ibaraki 305-0047, Japan
 }%
\author{Taro Fukazawa}%
 \affiliation{%
 CD-FMat, National Institute of Advanced Industrial Science and Technology, 1-1-1 Umezono, Tsukuba, Ibaraki 305-8568, Japan
 }%
 \affiliation{%
 ESICMM, National Institute for Materials Science, 1-2-1 Sengen, Tsukuba, Ibaraki 305-0047, Japan
 }%
\author{Guangzong Xing}%
\affiliation{%
ESICMM, National Institute for Materials Science, 1-2-1 Sengen, Tsukuba, Ibaraki 305-0047, Japan
}%
\author{Terumasa Tadano}%
 \affiliation{%
 CMSM, National Institute for Materials Science, 1-2-1 Sengen, Tsukuba, Ibaraki 305-0047, Japan
 }%
 \affiliation{%
 ESICMM, National Institute for Materials Science, 1-2-1 Sengen, Tsukuba, Ibaraki 305-0047, Japan
 }
\author{Takashi Miyake}%
 \affiliation{%
 CD-FMat, National Institute of Advanced Industrial Science and Technology, 1-1-1 Umezono, Tsukuba, Ibaraki 305-8568, Japan
 }%
 \affiliation{%
 ESICMM, National Institute for Materials Science, 1-2-1 Sengen, Tsukuba, Ibaraki 305-0047, Japan
 }%

\date{\today}% It is always \today, today,
%\date{September 10, 2005}

\begin{abstract}
%The maximum length is 600 characters including spaces.
Modern high-performance permanent magnets are made from alloys of rare earth and transition metal elements, and large magnetization is achieved in the alloys with high concentration of transition metals. 
We applied evolutionary search scheme based on first-principles calculations to the Y-Co-B system and predicted 37 cobalt-rich compounds with high probability of being stable. 
Focusing on remarkably cobalt-rich compounds, \ce{YCo_{16}} and \ce{YCo_{20}}, we found that, although they are metastable phases, the phase stability is increased with increase of temperature due to the contribution of vibrational entropy. 
The magnetization and Curie temperature are higher by 0.22\,T and 204\,K in \ce{YCo_{16}} and by 0.29\,T and 204\,K in \ce{YCo_{20}} than those of \ce{Y_{2}Co_{17}} which has been well studied as strong magnetic compounds. 
\end{abstract}

%\keywords{}

\maketitle

%\section{}
%\subsection{}
%\subsubsection{}

\section{Introduction}

Rare-earth magnets are strong permanent magnets, which mainly consist of rare-earth elements and 3d transition metals (Fe and/or Co). High Fe/Co concentration gives rise to high magnetization and rare earths are a source of high magnetocrystalline anisotropy which is essential for high coercivity. Rare-earth magnets have been developed since the discovery of large magnetocrystalline anisotropy in an alloy of yttrium and cobalt, \ce{YCo_{5}}~\cite{Hoffer1966}, and neodymium magnets are the strongest type of permanent magnet commercially available, and its main phase is formed by \ce{Nd_{2}Fe_{14}B} compound~\cite{Sagawa1984}, 
which has the saturation magnetization of 1.85\,T at 4.2\,K, magnetocrystalline
anisotropy field of 5.3\,MA/m at room temperature, and Curie
temperature of 586\,K~\cite{Hirosawa1986}. 

The magnetization is expected to be further increased using iron-richer compounds than \ce{Nd_{2}Fe_{14}B}, and  
\ce{$R$$T$_{12}} ($R$ = rare earth; $T$ = Fe, Co) systems have attracted considerable attention as potential candidates for stronger permanent magnets than \ce{Nd_{2}Fe_{14}B}~\cite{Ohashi1988-JApplPhys,Buschow1988,Yang1988}. 
It has long been known that \ce{$R$$T$_{12}} compounds are thermodynamically unstable in a bulk form and are stabilized by partial substitution of the third element for $T$,
\textit{i.e.} \ce{$R$($T$_{1-$x$}$X$_{$x$})_{12}} ($X$ = Al, Si, Ti, V, Cr, Nb, Mo, W)~\cite{Felner1983,Ohashi1987,Ohashi1988,DeMooij1988,Fuquana2005,Miyake2014,Harashima2016}. However, the magnetization decreases with the increase of $x$, and a search is currently underway for the best third elements, in other words, the elements maximizing the stabilization and minimizing the decrease of the magnetization. 
On the other hands, thin films of \ce{NdFe_{12}N_{$x$}} and \ce{Sm(Fe_{1-$x$}Co_{$x$})_{12}} have been fabricated by the epitaxial growth on W- and V-buffered \ce{MgO}(001) substrates. The films have higher magnetization, Curie temperature, and anisotropy field than \ce{Nd_{2}Fe_{14}B}~\cite{Hirayama2015,Hirayama2017}. 
In addition, \ce{YFe_{12}} with the \ce{ThMn_{12}} structure is experimentally obtained in multi-phases by rapid quenching method~\cite{Suzuki2017}. 
Recently, first-principles calculations predict that, in \ce{$Y$Fe_{12}} and \ce{Y(Fe_{1-$x$}Co_{$x$})_{12}} with $x$ of 0--0.7, the magnetization and Curie temperature are enhanced by the transformation from \ce{ThMn_{12}} into monoclinic $C2/m$ structures~\cite{Ishikawa2020-YFe12}.   

In the present study, we search for novel stable compounds with novel Fe/Co-rich rare-earth compounds using composition and crystal structure prediction scheme based on first-principles calculations and evolutionary algorithm. Here, we focused on Y-Co-B system for the following reasons: (i) Y is favorable for theoretical treatment because it has no $f$ electron in its ground-state electronic configuration, (ii) Co has a hexagonal close-packed (hcp) structure in the simple substance and is expected to be compatible with Y having hcp compared with Fe having a body-centered cubic structure, which makes a wide variety of compositions stable, and (iii) B can play a role for stabilizing Y-Co compounds and forming novel crystal structures, similarly to the case of \ce{Nd_{2}Fe_{14}B}. 
As a result, we found 37 cobalt-rich compounds including remarkably Co-rich \ce{YCo_{16}} and \ce{YCo_{20}}. 

\section{Computational details}

We used the evolutionary construction scheme of a formation-energy convex hull~\cite{Ishikawa2020-CH} to search for stable compounds in the Y-Co-B system. First, we created an initial set of  Y-Co-B compounds using the structure data of experimentally reported \ce{YFe_{3}}, \ce{Y_{6}Fe_{23}}, \ce{Y_{3}Fe_{29}}, \ce{NdFe_{2}}, \ce{NdFe_{5}}, \ce{Nd_{2}Fe_{17}}, \ce{Sm_{5}Fe_{19}}, \ce{YB_{2}}, \ce{YB_{4}}, \ce{CoB}, \ce{YCo_{2}B_{2}}, \ce{Y_{4}CoB_{13}}, \ce{Nd_{2}Fe_{14}B}, \ce{Sm_{2}Fe_{17}N_{3}}, and \ce{SmCo_{3}B_{2}}, which are included in the Materials Project database~\cite{Jain2013} and the SpringerMaterials database~\cite{SpringerMaterials}. 
%On the simple substances, 
For the pure elements, we used the hcp structures for Co and Y, and a rhombohedral $R\bar{3}m$ structure for B. 
Next, we constructed the preliminary convex hull of Y-Co-B system by performing the structural optimizations for the compounds in the initial set. 
Then, by applying evolutionary operators, ``mating'', ``mutation'', and ``adaptive mutation''~\cite{Ishikawa2020-CH} to target compounds selected from the compounds whose distance to the convex hull is small (0--4.4\,mRy/atom), new compositions and structures with high probability of being stable were created. Repeatedly performing the creation of compounds and the update of the convex hull, we searched for stable compounds. 

We combined our evolutionary construction code with the Quantum ESPRESSO (QE) code~\cite{QE} to perform the optimizations of the structures created by the operators. We used the generalized gradient approximation by 
Perdew, Burke and Ernzerhof (PBE) for the exchange-correlation functional in the framework of the projector augmented wave (PAW) method~\cite{PAW}. We got the PAW potentials from the QE web site~\cite{QEwebsite}. The energy cutoff was set at 100\,Ry for the wave function and 800\,Ry for the charge density. 
The maximum number of atoms in calculation cell is 84, and 
the $k$-space integration over the Brillouin zone (BZ) was carried out 
on 12 $\times$ 12 $\times$ 12, 8 $\times$ 8 $\times$ 8, 6 $\times$ 6 $\times$ 6, and 4 $\times$ 4 $\times$ 4 grids for the structures including 1--4, 5--12, 13--30, and more than 30 atoms in the calculation cell, respectively. 

To investigate dynamical and thermodynamical stability of the predicted compounds, we calculated phonon dispersion and vibrational free energy using the Vienna \textit{ab initio} simulation package (VASP)~\cite{vasp} and the PHONOPY code~\cite{phonopy}. 
Second-order interatomic force constants were computed by the finite-displacement method based on harmonic approximation, as implemented in PHONOPY. Total number of atoms in each supercell is $\sim$100 or larger, which was sufficient to reach the convergence of the vibrational free energy. 
The energy cutoff for the wave function was set at 400\,eV, and the $k$-point mesh was generated automatically in such a way that the mesh density in the reciprocal space is larger than 450 \AA$^{-3}$. The convergence criteria of energy and force minimization were set to 10$^{-8}$ and 10$^{-7}$ eV, respectively. 
%For structural optimization and phonon calculations, we used the Methfessel--Paxton smearing method~\cite{Methfessel_Paxton1989} with the width of 0.2 eV. On the other hand, the tetrahedron method with the Bl\"{o}chl correction~\cite{Blochl_PRB1994} was used for calculating the DFT static total energy.

%We estimated $T_{\text{C}}$ for the obtained structures within the mean-field approximation by the calculations of 
For the stable compounds, we calculated the intersite magnetic couplings using the Liechtenstein's method~\cite{Liechtenstein87}. 
For this purpose, we used AkaiKKR~\cite{AkaiKKR},
a first-principles program of Korringa-Kohn-Rostoker (KKR) Green's
function method, within the local density approximation.
%The number of k-points in the full Brillouin zone was
%14 $\times$ 10 $\times$ 10 for ThMn$_{12}$,
%12 $\times$ 12 $\times$ 8 for Type-I, and
%18 $\times$ 12 $\times$ 6 for Type-II. 
The Curie temperature $T_{\text{C}}$ is evaluated from a classical spin model within the mean-field approximation. 
%The other details of the estimation 
Other computational details are the same to the settings
in Ref.~\onlinecite{Fukazawa18}.

\section{Results}

\begin{figure}
\includegraphics[width=8.5cm]{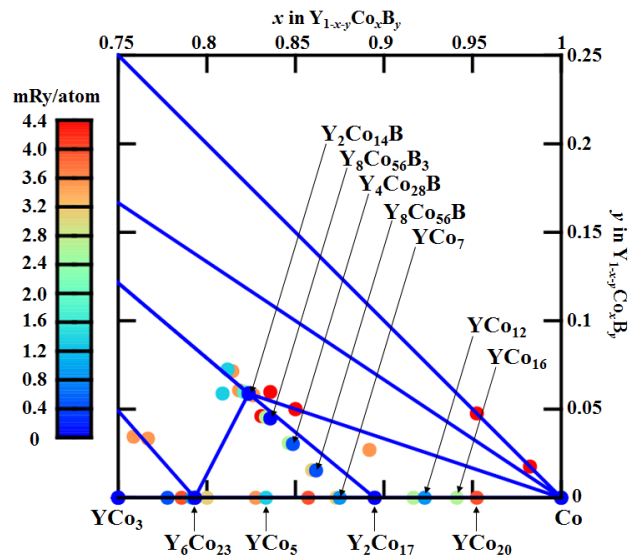}% Here is how to import EPS art
\caption{\label{Fig-convexhull} 
Projection of the formation-energy convex hull of \ce{Y_{$1-x-y$}Co_{$x$}B_{$y$}} on the $xy$ plane. The solid lines are the edges of the convex hull, and the dots show the compounds with the convex-hull distance less than 4.4\,mRy/atom.}
\end{figure}
In this study, we searched for stable and metastable compounds with the convex-hull distance $\Delta E$ less than 4.4\,mRy/atom (59.8\,meV/atom). 
This tolerance is associated with the approximations and the omission of temperature effects in first-principles calculations~\cite{Wu2013,Hinuma2016}, and the possibility of the stabilization by inclusion and/or substitution of other elements. 
We created 4120 structures up to the 11th generation by applying the evolutionary construction technique to the Y-Co-B system (\ce{Y_{$1-x-y$}Co_{$x$}B_{$y$}}, $0 \leq x \leq 1$, $0 \leq y \leq 1$) and predicted \ce{Y_{3}Co}, \ce{YCo}, \ce{YCo_{2}}, \ce{YCo_{3}}, \ce{Y_{6}Co_{23}}, \ce{Y_{2}Co_{17}}, \ce{YB_{2}}, \ce{YB_{3}}, \ce{CoB}, \ce{YCo_{2}B_{2}}, \ce{YCo_{3}B_{2}}, and \ce{Y_{2}Co_{14}B} as stable compounds on the convex hull (See Fig. S1 in Supplemental Material [SM]~\cite{SM_YCoB}). 
Figure \ref{Fig-convexhull} shows the closeup of the convex hull in the Co-rich region of $0.75 \leq x \leq 1$, in which we found four stable and 33 metastable compounds. 
The most important observation here is that \ce{YCo_{16}} and \ce{YCo_{20}} emerge as Co-richer compounds than \ce{YCo_{12}}. 
The $\Delta E$ values are 2.72\,mRy/atom for \ce{YCo_{16}} and 3.92\,mRy/atom for \ce{YCo_{20}}. 
%In Ref. \onlinecite{Wu2013}, the authors reported that more than 80\% of experimentally synthesized compounds included in the Inorganic Crystal Structure Database (ICSD) are stable or metastable compounds with $\Delta E$ less than 2.65\,mRy/atom in first-principles calculations. 
%Therefore, at least \ce{YCo_{16}} has a potential to be experimentally synthesized. 
In addition, for the metastable \ce{YCo_{5}} phase, we obtained an orthorhombic $Imma$ structure, which is more stable by 0.27\,mRy/atom than the \ce{CaCu_{5}}-type structure used to construct the preliminary convex hull (See Fig. S8 and Table S7 in SM for the details of the structure~\cite{SM_YCoB}). 
Another important point is that \ce{Y_{2}Co_{14}B} transforms into \ce{YCo_{7}} with an ordered tetragonal structure $P4_{2}/mnm$, going through the small energy region of $\Delta E$ less than 0.82\,mRy/atom. 
\ce{Y_{2}Co_{14}B} takes the \ce{Nd_{2}Fe_{14}B}-type structure with $P4_{2}/mnm$  including four formula units in the unit cell. The low-energy path connecting between \ce{Y_{2}Co_{14}B} and \ce{YCo_{7}} is achieved by step-by-step eliminating the B atoms from the unit cell (See Fig. S30 in SM~\cite{SM_YCoB}). 
See Fig. S2 in SM for the details of other metastable compounds~\cite{SM_YCoB}. 

\begin{figure}
\includegraphics[width=8.0cm]{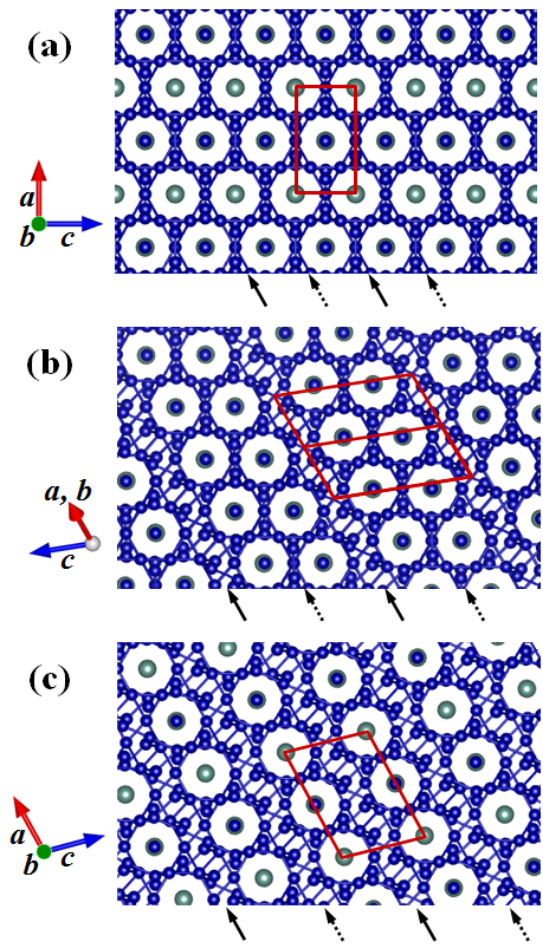}% Here is how to import EPS art
\caption{\label{Fig-structure} 
Projections of (a) \ce{YCo_{12}} with a tetragonal $I4/mmm$ (\ce{ThMn_{12}}-type) structure along the $b$ axis, (b) \ce{YCo_{16}} with a triclinic $P\bar{1}$ structure along the $[1\bar{1}0]$ direction, and (c) \ce{YCo_{20}} with a monoclinic $C2/m$ structure along the $b$ axis. Frames show the unit cells, and large and small balls represent the Y and Co atoms, respectively. The solid (broken) arrows show the areas where additional Co atoms are inserted by the transformation from \ce{YCo_{12}} (\ce{YCo_{16}}) into \ce{YCo_{16}} (\ce{YCo_{20}}). Crystal structures were drawn with VESTA~\cite{VESTA}.}
\end{figure}
Hereafter, we focus on the novel Co-rich compounds, \ce{YCo_{16}} and \ce{YCo_{20}}. 
The structures are assigned as triclinic $P\bar{1}$ for \ce{YCo_{16}} and monoclinic $C2/m$ for \ce{YCo_{20}} (See Tables S27 and S28 and Fig. S28 and S29 in SM for the details of the structures~\cite{SM_YCoB}). 
Figure \ref{Fig-structure} shows $I4/mmm$ (\ce{ThMn_{12}}-type) \ce{YCo_{12}} viewed along the $b$ axis, $P\bar{1}$ \ce{YCo_{16}} viewed along the $[1\bar{1}0]$ direction, and $C2/m$ \ce{YCo_{20}} viewed along the $b$ axis. 
$P\bar{1}$ \ce{YCo_{16}} and $C2/m$ \ce{YCo_{20}} are achieved by adding the Co atoms into \ce{ThMn_{12}}-type \ce{YCo_{12}}. 
\ce{YCo_{16}} (\ce{YCo_{20}}) is obtained by the insertion of the four Co atoms per formula unit into the area shown by the solid (broken) arrows of \ce{YCo_{12}} (\ce{YCo_{16}}), parallel to the (101) plane of \ce{ThMn_{12}}-type \ce{YCo_{12}}. 

\begin{figure}
\includegraphics[width=8.2cm]{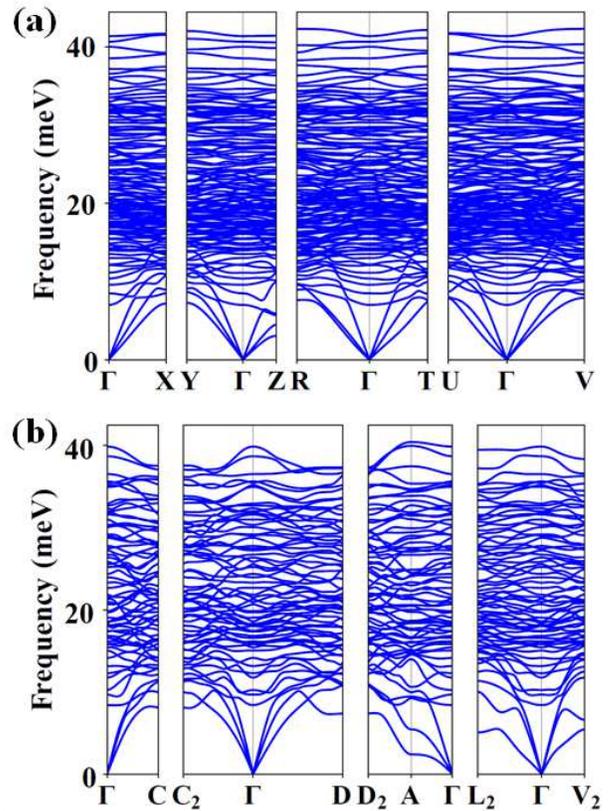}% Here is how to import EPS art
\caption{\label{Fig-phonon} 
Phonon dispersions of (a) \ce{YCo_{16}} with a triclinic $P$-1 structure and (b) \ce{YCo_{20}} with a monoclinic $C2/m$ structure.}
\end{figure}
\begin{figure}
\includegraphics[width=8.0cm]{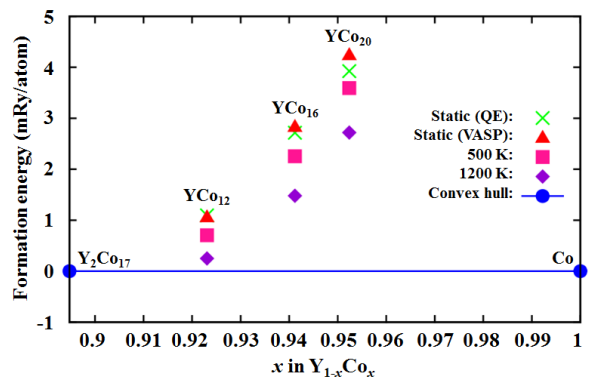}% Here is how to import EPS art
\caption{\label{Fig-temperature} 
Phase stability of \ce{YCo_{12}}, \ce{YCo_{16}}, and \ce{YCo_{20}} against the decomposition into \ce{Y_{2}Co_{17}} and Co.}
\end{figure}
We investigated the dynamical and thermodynamical stability of \ce{YCo_{16}} and \ce{YCo_{20}} by performing phonon calculations. 
Figure \ref{Fig-phonon} shows the phonon dispersion curves of $P\bar{1}$ \ce{YCo_{16}} and $C2/m$ \ce{YCo_{20}}. 
No imaginary phonon modes were detected in the dispersion curves, which indicates that the two structures are dynamically stable at 0\,K. 
We investigated the variations of the convex-hull distance for \ce{YCo_{16}} and \ce{YCo_{20}} with increase of temperature by considering the entropy contribution, including electronic and vibrational free energies (Fig. \ref{Fig-temperature}). 
We compared the static formation energies of $I4/mmm$ \ce{YCo_{12}}, $P\bar{1}$ \ce{YCo_{16}}, and $C2/m$ \ce{YCo_{20}} between QE and VASP and confirmed the errors are 0.03\,mRy/atom, 0.12\,mRy/atom, and 0.32\,mRy/atom, respectively. Since these results are reasonably consistent with each other, we discuss the finite-temperature thermodynamic stability of \ce{YCo_{16}} and \ce{YCo_{20}} based on the VASP results. %These results suggest that the thermodynamic stability of \ce{YCo_{16}} and \ce{YCo_{20}} is quantitatively discussed with the use of VASP. 
With increasing the temperature up to 1200\,K,
the convex-hull distances decrease to 0.25\,mRy/atom for \ce{YCo_{12}}, 1.48\,mRy/atom for \ce{YCo_{16}}, and 2.72\,mRy/atom for \ce{YCo_{20}}. 
%We note that, although \ce{Y_{2}Co_{17}} has a rhombohedral $R\bar{3}m$ (\ce{Th_{2}Zn_{17}}-type) structure at low temperatures and a hexagonal $P6_{3}/mmc$ (\ce{Th_{2}Ni_{17}}-type) structure at 1200\,K due to the entropy contribution. 
Regarding the structure of \ce{Y_{2}Co_{17}}, a rhombohedral $R\bar{3}m$ (\ce{Th_{2}Zn_{17}}-type) structure is stable in the low-temperature region, while the entropy contributions make the hexagonal $P6_{3}/mmc$ (\ce{Th_{2}Ni_{17}}-type) more stable at temperatures above 780\,K. This temperature-induced phase transition was properly considered in the results shown in Fig.~\ref{Fig-temperature}. 

\begin{table}
\caption{\label{data_comparison}
Comparison of volume per atom $V$, magnetic moment per atom $m$, magnetization $M$, and Curie temperature $T_{\text{C}}$ among \ce{Y_{2}Co_{17}}, \ce{YCo_{12}}, \ce{YCo_{16}}, \ce{YCo_{20}}, \ce{Y_{2}Fe_{17}}, \ce{YFe_{12}}, \ce{YFe_{16}}, and \ce{YFe_{20}}.}
%\begin{center}
\begin{ruledtabular}
\begin{tabular}{cccccc}
& structure & $V$ & $m$ & $M$ & $T_{\text{C}}$ \\
& & (\AA$^{3}$/atom) & ($\mu_{\text{B}}$/atom) & (T) & (K) \\ \hline
\ce{Y_{2}Co_{17}} & $R\bar{3}m$ & 12.60 & 1.354 & 1.252 & 1174\\
\ce{YCo_{12}} & $I4/mmm$ & 12.15 & 1.455 & 1.396 & 1280\\
\ce{YCo_{16}} & $P\bar{1}$ &11.90 & 1.505 & 1.474 & 1378\\
\ce{YCo_{20}} & $C2/m$ & 11.77 & 1.555 & 1.539 & 1378\\ \hline
\ce{Y_{2}Fe_{17}} & $R\bar{3}m$ & 13.42 & 1.979 & 1.719 & 720\\
\ce{YFe_{12}} & $I4/mmm$ & 12.88 & 2.019 & 1.826 & 792 \\
\ce{YFe_{16}} & $P\bar{1}$ & 12.65 & 2.093 & 1.928 & 719 \\
\ce{YFe_{20}} & $C2/m$ & 12.41 & 2.053 & 1.928 & 434 \\
\end{tabular}
%\end{center}
\end{ruledtabular}
\end{table}
Next, we investigated the magnetic properties of $P\bar{1}$ \ce{YCo_{16}} and $C2/m$ \ce{YCo_{20}}. Table \ref{data_comparison} lists volume per atom $V$, magnetic moment per atom $m$, total magnetization $M$, and Curie temperature $T_{\text{C}}$ for \ce{Y_{2}Co_{17}, }\ce{YCo_{12}}, \ce{YCo_{16}}, and \ce{YCo_{20}}. The $V$ and $m$ values decrease and increase with the increase of the Co concentration, respectively. 
Consequently, the $M$ value increases to 1.474\,T in \ce{YCo_{16}} and 1.539\,T in \ce{YCo_{20}}, which are larger than those in \ce{Y_{2}Co_{17}} and \ce{YCo_{12}}. 
Furthermore, we obtained that \ce{YCo_{16}} and \ce{YCo_{20}} show the $T_{\text{C}}$ value of 1378\,K, which is higher by 204\,K and 98\,K than those of \ce{Y_{2}Co_{17}} and \ce{YCo_{12}}, respectively. 
Although the mean-field approximation tends to overestimate $T_{\text{C}}$, the differences of theoretical $T_{\text{C}}$ values among magnet compounds have been found to be qualitatively consistent with those in experiments~\cite{Fukazawa17,Fukazawa18,Fukazawa2019-Tc}. Therefore, we expect the trend of the enhancement in $T_{\text C}$ is realistic.

We also calculated the $V$, $m$, $M$, and $T_{\text{C}}$ values for Fe-based compounds. Although the $m$ value increases with increase of the Fe concentration from \ce{Y_{2}Fe_{17}} through \ce{YFe_{16}}, it turns to the decrease in \ce{YFe_{20}} due to the decrease of the magnetic moment of Fe at the 4$e$ site. As a result, the $M$ value shows the increase to 1.928\,T in \ce{YFe_{16}}, whereas there is no further increase in \ce{YFe_{20}}. In contrast to the case of the Co-based compounds, the $T_{\text{C}}$ value decreases to 719\,K in \ce{YFe_{16}}, which is almost equal to that of \ce{Y_{2}Fe_{17}}, and is largely decreased in \ce{YFe_{20}}. 
%These results suggest that, in the Fe-based systems, \ce{YFe_{16}} is a better candidate for high-performance magnets than \ce{YFe_{20}}. 

\section{Conclusion}

In conclusion, we searched for stable compounds in the Y-Co-B system, \ce{Y_{$1-x-y$}Co_{$x$}B_{$y$}}, using the evolutionary construction technique of a formation-energy convex hull. Focusing on Co-rich ($0.75 \leq x \leq 1$) and low energy ($\Delta H \leq 4.4$\,mRy/atom) region, we predicted 34 compounds, including novel Co-rich compounds, \ce{YCo_{16}} and \ce{YCo_{20}}. In addition, we obtained a new stable structure of \ce{YCo_{5}}, and a low energy path connecting between \ce{Y_{2}Co_{14}B} and \ce{YCo_{7}}. Phonon calculations predict that \ce{YCo_{16}} and \ce{YCo_{20}} are dynamically stable and the hull distance $\Delta H$ is decreased to 1.48\,mRy/atom for \ce{YCo_{16}} and 2.72\,mRy/atom for \ce{YCo_{20}} with increase of temperature to 1200\,K due to the contribution of the vibrational free energy. 
The calculated $M$ and $T_{\text{C}}$ values are 1.474\,T and 1378\,K in \ce{YCo_{16}} and 1.539\,T and 1378\,K in \ce{YCo_{20}}, which are larger than those in \ce{Y_{2}Co_{17}} and \ce{YCo_{12}}. 
%These results suggest that \ce{YCo_{16}} and \ce{YCo_{20}} are expected to be novel high-performance permanent magnets. 
We performed the same calculations for \ce{YFe_{16}} and \ce{YFe_{20}}, and obtained the results that, although the $T_{\text{C}}$ values are decreased in \ce{YFe_{16}} and \ce{YFe_{20}}, the $M$ values are increased. 

This study provides novel Co-rich compounds, \ce{YCo_{16}} and \ce{YCo_{20}}, with high magnetization and high Curie temperature, whereas further studies are required to get the knowledge about their magnetocrystalline anisotropy and coercivity, which is crucial for the application of \ce{$R$$T$_{16}} and \ce{$R$$T$_{20}} systems to high-performance permanent magnet. 
For example, it is important to accumulate the data about the variation of the magnetocrystalline anisotropy and coercivity by systematically replacing Y with the other $R$ elements. 

\begin{acknowledgments}
This work was supported by the Ministry of Education, Culture, Sports, Science and Technology (MEXT) as ``The Elements Strategy Initiative Center for Magnetic Materials (ESICMM)'' (JPMXP0112101004) and ``Program for Promoting Researches on the Supercomputer Fugaku'' (DPMSD).
The computation was partly conducted using the facilities of the Supercomputer Center, the Institute for Solid State Physics, the University of Tokyo, the supercomputer of ACCMS, Kyoto University, and the HPCI System Research project (Project ID:hp200125).
\end{acknowledgments}
%\bibliography{References}

\end{document}